# Digital Twin in Practice: Emergent Insights from an ethnographic-action research study

Ashwin Agrawal, M.S.[1]; Vishal Singh, Ph.D.[2]; Robert Thiel, M.S.[3];
Michael Pillsbury, M.A.[4]; Harrison Knoll, M.S.[5]; Jay Puckett, M.A.[6] ; Martin Fischer, Ph.D.[7]

[1]Department of Civil and Environmental Engineering, Stanford University, Stanford, CA , USA (Corresponding author) Email: ashwin15@stanford.edu
[2]Associate Professor, Centre of Product Design and Manufacturing, Indian Institute of Science, Bangalore, India, Email: singhv@iisc.ac.in
[3]Vice President, WSP USA, Morristown, New Jersey, USA, Email: bob.thiel@wsp.com
[4]Partner, Henniker Associates LLC, USA, Email: mpillsbury@louisberger.com
[5]Chief Executive Officer, ROCK robotic, Email: harrison.knoll@gmail.com
[6]Associate Vice President, WSP USA, Morristown, New Jersey, USA, Email: jay.puckett@wsp.com
[7]Professor, Civil and Environmental Engineering, Stanford University, Stanford, CA, USA Email: fischer@stanford.edu

## ABSTRACT

Based on an ethnographic action research study for a Digital Twin (DT) deployment on an automated highway maintenance project, this paper reports some of the stumbling blocks that practitioners face while deploying a DT in practice. At the outset, the scope of the case study was broadly defined in terms of digitalization, and software development and deployment, which later pivoted towards the concept of Digital Twin during the collective reflection sessions between the project participants. Through an iterative learning cycle via discussions among the various project stakeholders, the case study led to uncovering the roadblocks in practice faced by the Architecture, Engineering, and Construction (AEC) practitioners. This research finds that the practitioners are facing difficulty in: (1) Creating a shared understanding due to the lack of consensus on the Digital Twin concept, (2) Adapting and investing in Digital Twin due to inability to exhaustively evaluate and select the appropriate capabilities in a Digital Twin, and (3) Allocation of resources for Digital Twin development due to the inability to assess the impact of DT on the organizational conditions and processes.

## INTRODUCTION

The rise of digitalization over the last couple of decades has opened new avenues for the construction industry in terms of value creation (Hampson and Tatum 1993). By providing bi-directional coordination between the virtual (digital) and the physical worlds, Digital Twin (DT) has emerged as a key concept in the digital transformation of the construction industry. It offers opportunities for dramatic improvements in effectiveness and efficiency through visualization, information integration, and automation.

Although literature reports about the specific purposes and various task-related benefits of DT (e.g., Khajavi et al. 2019; Opoku et al. 2021), little is known about the stumbling blocks that practitioners face while deploying a DT in practice. A few similar studies uncovering pragmatic insights from a DT deployment (e.g., Feng et al. 2020) have been conducted in the manufacturing industry, but nothing much exists in the AEC industry. One reason for this might be that actual DT development and deployment happens over several months, if not years, making it difficult to conduct a longitudinal study for understanding the roadblocks in the practical deployment of a DT. Furthermore, the limited interest of practitioners to be willing to collaborate and give access to information on such a large time scale makes it even more difficult.

This paper reports the results from an ethnographic action research study on a real-life DT deployment in the AEC industry. The article is built around the findings and observations from an applied case study that started with a clear set of digitalization goals and objectives. The study started as a digitalization implementation on a real-life project in which the clients were looking for an automated assessment of the health condition of the road surface and the surrounding areas of an active tollway. As the project progressed, discussions emerged about DT between the practitioners as they reflected on how it related to the work the researcher was doing. These discussions raised some interesting questions and made the researcher wonder what are the problems that practitioners face while deploying a DT in practice.

The study uses ethnographic-action research methodology (Hartmann, Fischer, and Haymaker 2009) to report the findings of the case study. Ethnographic-action research combines the implementation of a system in practice with ethnographic observations of the practitioners. This type of research methodology is particularly relevant in the construction management domain, where practical problems need to be solved simultaneously creating theoretical knowledge (Azhar, Ahmad, and Sein 2010). Ethnographic-action research has been popular for investigating the use of technology in construction projects (Oswald and Dainty 2020; Zanen et al. 2013; Mahalingam, Yadav, and Varaprasad 2015). In our study, results are reported based on 10-month longitudinal data, collected between Autumn 2018 and Summer 2019, consisting of participant observations, informal discussions, meeting observations, document analysis, and semi-structured interviews with various stakeholders at different levels of management. A site visit was also conducted, permitting the research team to interact with on-field personnel, project managers, and executives from the Department of Transportation (DoT). In the next sections, the paper describes the DT and the case study background. This is followed by reporting the observations and questions arising from the case study. The paper is concluded with the discussions of the findings and their implications for the AEC industry.

**DIGITAL TWIN BACKGROUND**

While the term DT has been coined recently, the concept was first introduced as 'Mirrored Spaces Model' by Michael Grieves in 2002 as a part of a university course on Product Lifecycle Management (Grieves and Vickers 2017). DT essentially consists of four major elements: (1) The physical object, (2) The virtual or the digital object, (3) Link for data flow from physical to virtual object, and (4) Link for data flow from virtual to physical object. The data flow between the two systems helps in mirroring or twinning of what exists in the real world to the virtual world and vice-versa. The first formal definition of the DT was formulated by NASA in 2012 (Glaessgen and Stargel 2012) as: "an integrated multi-physics, multi-scale, probabilistic simulation of a vehicle or system that uses the best available physical models, sensor updates, fleet history, etc., to mirror the life of its flying twin". Thereafter, owing to its usefulness, the concept has spread from the aerospace sector to a wider range of industries. In the manufacturing industry, DT has been used for applications including Product Lifecycle Management (Boschert and Rosen 2016), Production Planning and Control (Söderberg et al. 2017), process redesign (Schleich et al. 2017), and layout planning (Rosen et al. 2015). In the AEC industry, DT has been used in building operations and maintenance (Khajavi et al. 2019), logistics processes, and energy simulations (Opoku et al. 2021).

**CASE STUDY BACKGROUND**

The case study is focused on the 41-mile southern section (Section 5 & 6) of SH-130, a publicly owned toll road, privately operated and maintained by the SH-130 concession company. Opened in 2012, the $1.35 billion facility was built and financed by the SH-130 concession company and is operated and maintained under the terms of a 50-year facility concession agreement with the Texas Department of Transportation (TxDoT). Interestingly, SH-130 segments 5 & 6 have the highest legal speed limit in the nation at 85 mph and uses open tolling to allow tolls to be charged without drivers having to slow down for a toll booth. To ensure a safe driving experience for the users and sustain such high speeds, timely maintenance of the roadway is required. Therefore, the focus of the concession company during the case study was to come up with a strategy to improve the efficiency of the costliest phase of the highway lifecycle: operations and maintenance.

The developer and the operator of the road have a general obligation to maintain and operate the facility in accordance with good industry practice. This includes minimizing the delays for the users, responding as quickly as possible to incidents and defects, and protecting the safety of users in and around the facility. The performance requirements were set upfront in the SH-130 concession agreement by the Department of Transportation (DoT), indicating the minimum expected performance levels and the defect resolution timing and procedures. For some requirements, a very stringent response time - 6 to 24 hours - was specified in the concession. Failing to respond adequately would result in heavy fines and risks to the life safety of the users. These requirements necessitate a significant investment into inspection and maintenance management, including both the personnel time and resources. Therefore, the company wondered if digitalizing some of the ways of working can help.

Specifically, the company was interested in investigating novel methods of data acquisition and data analysis to develop a preventative maintenance program to reduce the TCO (total cost of ownership) through reduced operational costs. The vision that the company had for the project was: "Reduction of operational costs through early detection and preventative maintenance." New data acquisition methods from unmanned aerial imagery, mobile imagery, mobile LiDAR, and unmanned LiDAR systems were explored, and machine learning techniques for automated data analysis were applied. Some example images taken during the project are shown in Figure-1. The authors took the lead on developing the algorithms to process real-time data of the roadway through UAV systems, and analyze it using machine learning. The resulting prototype was evaluated to understand the value it generates for the company. The prototype application as a result of digitalization on the project started to be called "DT for the roadway" by the practitioners and thereafter became the center point of discussion in the project. The project concluded with a successful proof of concept of the DT showcasing that the commercially available technology was able to identify critical performance requirements from the concession agreement. To observe the evolution of the understanding of the DT concept by the practitioners, and different stakeholders, notes were made throughout the project, and minutes of meetings were recorded.

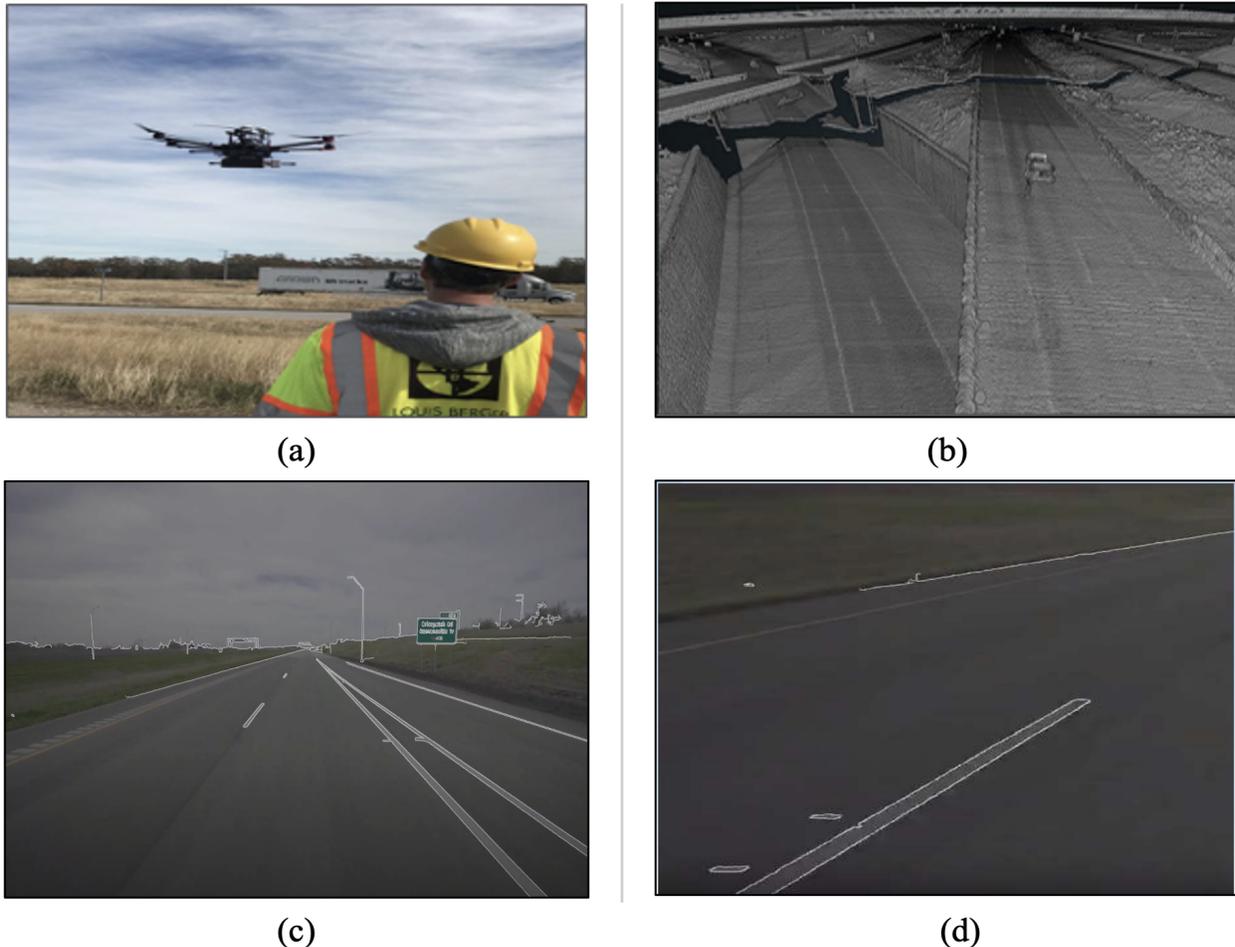

*Figure 1: Photos taken during data collection and analysis - (a) Photograph taken during data collection while flying the drone over SH-130; (b) Drone image of SH-130 with bridge; (c) Computer vision algorithm detecting the road elements near the pavement in an image taken from a camera mounted on a car; (d) Computer vision algorithm detecting road markings in an image taken by the drone.*

**CASE STUDY FINDINGS**

*Insight-1: Lack of consensus on the DT concept among the practitioners*

A lack of consensus among the practitioners about the DT concept was observed. For some people, the term DT implied an entirely new way of doing things with technology. And for others, DT was just an extension of the already existing technology like Building Information Modelling (BIM).

*Supporting Observations:*

(1) The researchers observed that the practitioners found it difficult to semantically differentiate between the technologies: BIM, DT, BIM coupled with Internet of Things (IoT) [BIM + IoT], etc., and sometimes used the terms interchangeably. Some excerpts from the practitioners like: "This is like a BIM for horizontal projects", "We can use sensors on the road to update the model", etc. were noted in the case study. This ambiguity in the DT concept and lack of its ability to differentiate from general computing models has also been suggested by (Shao and Helu 2020; Khajavi et al. 2019) in the manufacturing industry.

(2) DT wasn't even an explicit goal that the project team was initially pursuing and emerged more as a discussion point under the broader purview of digitalization. After the introduction of the term, one of the practitioners stated: "Yes, exactly we are building a DT of the road". In the minds of the practitioners, they were already building a DT, and the formal introduction of the term DT gave them the means to communicate their prototype to a broader audience. The researchers observed that practitioners were unaware of the technicalities of the term DT, and were indifferent if the task was executed by a BIM or a DT.

*Implications:*

The usage of DT across such a wide range of things and applications risks it being rejected by the people due to its vagueness and as hype (Wright and Davidson 2020). In order to have a clear understanding of the concept of DT, a way to describe, categorize and compare the different perspectives is needed. The authors believe that articulating the DT concept more clearly and defining "Levels of Digital Twin" (Madni, Madni, and Lucero 2019) can be one of the ways that can help to create a hierarchy/categorization of perspectives and thus help to compare the different types of DT.

*Insight-2: Inability to exhaustively evaluate the capabilities of a DT*

Even though the ambition in many cases is to create the so-called "Most intelligent DT", that can completely automate the work processes, the reality is that in the current practice and the foreseeable future, DT would still have limited technological capabilities. Therefore, it is important for the practitioners to be able to select the appropriate technological capabilities by exhaustively evaluating the various alternatives in a DT. Towards this end, the researchers observed that practitioners are facing difficulty in selecting the appropriate technological capabilities in a DT while deploying it in practice.

*Supporting Observations:*

It was highlighted in one of the semi-structured interviews that the whole DT prototyping effort was top-down driven. The decision to use drone imagery and machine learning was more of a top management fascination with the drone technology, rather than an exhaustive evaluation of all possible alternative options to detect the defects like using some sensor-based technologies, monitoring change in speed of vehicles, etc. Similar observations highlighting the importance of a clear scope definition and exhaustive evaluation have also been made in the other domains, where (Feng et al. 2020; Shao and Helu 2020) argues that the idea that Digital Twin (DT) has to be a comprehensive virtual replica of all the physical assets along with increasingly complex capabilities like Artificial Intelligence (AI) and Machine Learning (ML) is making it difficult for the practitioners to adapt, invest, and deploy DT.

*Implications:*

The lack of an exhaustive evaluation of alternative options can lead to biases towards a particular technology and therefore result in missed opportunities. A critical evaluation of the technology, and questioning 'how' digital technologies can coalesce to generate business value and improve business competitiveness is needed (Love and Matthews 2019). A standardized framework providing guidelines, methods, and best practices, can enable practitioners to leverage DT by providing a means to navigate the complex set of standards, technologies, and procedures that can be used for DT implementation (Shao and Helu 2020).

### *Insight-3: Inability to assess the impact of DT on the organizational conditions and processes*

When the discussions shifted towards deploying the DT, the team realized that in order to get the benefits from the DT, they would need to change the current ways of working. The practitioners discussed various possible changes that might result due to the DT. Although the practitioners brainstormed on the possible impacts, a structured discussion involving clear communication of ideas was missing.

*Supporting Observations:*

(1) The practitioners discussed that the process of creating a work order for a damage repair, updating the digital asset model for some observed defects, closing a work order after inspection, etc. can all be automated as the drone would be able to detect the defects and understand the changes that happen over time. The practitioners, therefore, wondered how would this affect the current ways of working. Some ideas that propped up included shifting the personnel dedicated to these tasks for more productive tasks.

(2) The practitioners also reflected that owing to the stringent guidelines for the road repair, currently, the SH-130 concession company has to keep a large amount of inventory and contact individual subcontractors in a short period of time to carry out the repair work leading to unnecessary costs for the company. Once the team moves towards a predictive maintenance system using drones, they can start planning in advance for the material, crew, equipment, etc. This would allow the company to pivot to a "Just-in-time" delivery method for materials and equipment delivery, and encourage competitive bids among the subcontractors. Again many plausible impacts were suggested in form of educated guesses without a definite structure like changing the existing contracts, legal tenders, business partnerships, etc.

*Implications:*

The benefits of DT are marginal if only superimposed on the existing construction management processes and organizational structures (Venkatraman 1994). The maximum benefit accrues when investments in DT are accompanied by the corresponding changes in the process and organization delivering the product/service. Therefore, a lack of structured impact assessment and change management can lead to implementation lags and misallocation of resources. Possibly, methods similar to Benefits Dependency Network (Love and Matthews 2019) which provide a structured way to assess the changes required to sustain the technology can be used in the context of DT to alleviate this challenge.

**CONCLUSION**

Digitalization approaches are pragmatically determined largely by how it gets interpreted by the practitioners and applied in practice. Therefore, understanding the roadblocks occurring in practice becomes critical to lay out the guidelines for the implementation of DT in the AEC industry and future academic research. Based on the ethnographic action study, three main roadblocks have been identified and presented, with the hope, the many other similar additional insights emerge from future studies.

We find that there is a lack of consensus on the DT concept in practice. As these concepts are critical for creating a shared understanding and communication of the digitalization goals among the project participants, a clear articulation of the DT definition and a framework for unifying the various perspectives is needed. We also find that practitioners face difficulty in exhaustively evaluating the varied capabilities of DT, and thus struggle with deciding the appropriate scope for deployment leading to difficulty in adaption and missed opportunities. Finally, we also note that practitioners struggle to articulate the consequent changes that need to happen in the organization and its processes to sustain the generated value from the DT.